\documentstyle[preprint,aps,prb,epsf,psfig]{revtex}
\begin{document}

\title{ Origin of the unusual dependence of Raman D band on
excitation wavelength in graphite-like materials
}
\author{A. K. Sood, Rajeev Gupta and S. A. Asher$^{*}$
\\
}
\address{Department of Physics, Indian Institute of Science,
Bangalore, India 560 012
}
\address{$^{*}$
Department of Chemistry, University of Pittsburgh, Pittsburgh,
PA 15260, USA}
\maketitle

\begin{abstract}
We have revisited the still unresolved puzzle of the dispersion of
the Raman disordered-induced D band as a function of laser excitation photon energy E$_L$ in 
graphite-like materials.  We propose that the D-mode is a combination 
of an optic phonon at the K-point in the Brillioun zone and an acoustic 
phonon whose momentum is determined uniquely by the double resonance condition.
The fit of the experimental data with the double-resonance model yields 
the reduced effective mass of 0.025m$_{e}$ for the electron-hole pairs 
corresponding to the A$_{2}$ transition, in agreement with other experiments. The model 
can also explain the difference between $\omega_S$ and $\omega_{AS}$ for D and D$^{\star}$
 modes, and predicts its dependence on the Raman excitation frequency.

\noindent{PACS numbers: 63.20.Dj, 63.50.+x, 78.30.Ly \\}
\end{abstract}

\section{Introduction}
Carbon based materials, ranging from highly oriented pyrolitic
graphite, diamond like carbon films, fullerenes and carbon nanotubes 
have been the subjects of extensive studies over the past two 
decades because of
their numerous technological applications.
Raman spectroscopy is a powerful non-destructive probe for
characterising different forms of carbon materials-diamond,
crystalline graphite, glassy carbon, hydrogenated amorphous carbon,
fullerenes and nanotubes. Raman spectra of these materials gives
considerable information on the nature of chemical bonding (sp$^2$
vis-a-vis sp$^3$) and  relative
abundance of sp$^2$ and sp$^3$ bonds, residual stresses, crystallite
size and dopants. There is a long-standing puzzle in the Raman spectra of
graphitic materials. Single crystals of pristine graphite
($D_{6h}^{4}$ space group symmetry) has two in-plane Raman active
E$_{g}$ modes: one at 42 cm$^{-1}$ and the other at $\sim$ 1582 cm$^{-1}$
(G-band)\cite{c8tuins1}. In presence of disorder, an additional Raman band 
(called D-band) is observed at $\sim$ 1350 cm$^{-1}$ for a laser excitation energy 
of 2.41 eV (wavelength  
514.5 nm), irrespective of the kind of carbon material \cite{c8tuins1,c8wang1,c8thomsen1,c8pocsik1}.

The important features of the
Raman spectra associated with the disorder-induced D-band are as
follows: (i) The
E$_{L}$-dependence of the D-band frequency is essentially independent
of the type of carbon involved. It occurs in all sp$^{2}$-
hybridized disordered carbon materials, powdered and randomly oriented
crystalline graphite, glassy carbon, boron-doped highly ordered
pyrolytic graphite, carbon black, multicomponent carbon films,
hydrogenated amorphous carbon and more recently, in carbon nanotubes \cite{c8thomsen1}.
(ii) The frequency of the D-band shifts upward with increasing
excitation laser energy \cite{c8vidano1} E$_{L}$ : $\omega_{D}$ =
1279 cm$^{-1}$ for E$_{L}$ = 1.17 eV and $\omega_{D} \sim$  1410
cm$^{-1}$ for E$_{L}$ = 3.54 eV ($\simeq$ 40 cm$^{-1}$/eV). Fig.1 shows the data from various
measurements made on various kinds of disordered carbon \cite{c8wang1,c8pocsik1} and carbon
nanotubes \cite{c8thomsen1}. (iii) The frequency of the associated
second order (D$^{\star}$) band ($\sim$ 2700 cm$^{-1}$ for E$_{L}$=2.41 eV)
also shifts up with increasing  E$_{L}$. The dispersion
$d\omega_{D^{\star}}/dE_{L}$  is almost twice that of
$d\omega_{D}/dE_{L}$. (iv) The E$_{L}$-dependant second order
D$^{\star}$ band is observed in Raman spectra of single crystal graphite even
though the first order disorder induced D-mode is absent \cite{c8vidano1}. (v) The D-band intensity decreases as E$_{L}$ increases from 1.16 eV
to 4 eV. In contrast, the intensity of the G-band is maximum at
E$_{L}$ $\sim$5 eV \cite{c8sood1}. The D band is not observed with excitation
energies E$_{L} >$ 4.2 eV as shown in Fig.2 which displays Raman spectra of graphite powder (with a small amount of sodium sulphate for internal intensity calibration) using E$_{L}$ = 2.54 eV and 4.82 eV. The lines marked by asteriks are from internal vibrational modes of sulphate ions. The intensities have been normalised with respect to the sulphate mode at 980 cm$^{-1}$. The data in the inset taken from Wang {\it et al.} \cite{c8wang1} show the intensity ratio of D band with respect to that of G band for 1.16 $< E_{L} < 4$ eV. The decrease of this intensity ratio as E$_{L}$ increases has been seen by others as well \cite{c8pocsik1}.
In other words, the resonance excitation profile of the D-band will
show a maximum at  E$_{L}$ less than 1.16 eV, which is very different from the
G-band behaviour. These experimental observations are crucial to arrive at the correct explanation for the origin of the D-band. (vi) The peak positions of the D and D$^{\star}$ bands
observed in Stokes ($\omega_{s}$) and anti-Stokes are not the same
\cite{c8ptan1}: $\Delta\omega_{S}=\omega_{AS}-\omega_{S}$ =7 cm$^{-1}$
for the D-mode and  $\omega_{AS}-\omega_{S}$ =28 cm$^{-1}$ for the D$^{\star}$
mode when E$_{L}$ =2.41 eV. In order to make sure that this anomalous
difference between Stokes and anti-Stokes peak positions of the D and
D$^{\star}$ bands is genuine, it was seen in the same experiment that  $\Delta\omega
\sim$ 0 for the G-band, as is usually expected.

It is surprising that even after twenty years of experimental observation
\cite{c8tuins1}, the origin and dispersion of the D-band is still not
quantitatively understood. Tuinstra and Koenig \cite{c8tuins1} attributed the
D-mode to the A$_{1g}$ type mode of a finite graphite crystallite with
symmetry lower than D$_{6h}$  of graphite, which corresponds to a
longitudinal acoustic mode for the infinite lattice. Nemanich and
Solin \cite{c8neman1} and Lespande {\it et al.}\cite{c8lesp1} have
invoked the breakdown of wave-vector selection rules due to disorder
or finite size of crystals and correlated the D-band features with the
density of states of graphite. These models, however, cannot explain the
strong dispersive nature of the D-mode as a function of E$_{L}$. The observation of
second-order D$^{\star}$ band for single crystal graphite which shows no
disorder-induced D mode also rules out the explanation of the
dispersion to crystallite size selective resonance Raman scattering
process which occurs in polyacetylene. The recent studies \cite{c8mathew1,c8pocsik1} explained the 
dispersion of the D-mode following Boronov {\it et
al.}\cite{c8boron1} who attributed  the dispersion of
the D-band  in terms of the coupling between electrons and phonons
with the same wave-vector near the K-point of the Brillouin zone. This
model has a serious drawback that it does not justify the reasons for the coupling and cannot address the resonance of the D-mode at $E_{L}
<  1 eV$, as well as the difference in $\omega_{AS}$ and $\omega_{S}$ for the D
and D$^{\star}$ modes.

Almost three years back \cite{c8sood1}, we have proposed disorder-induced double resonance near a gap of $\sim$  1 eV in the graphite band structure to result in the dependence of the phonon wavevector q and hence the phonon frequency $\omega$ on the energy of the exciting laser energy. Very recently, Thomsen and Reich \cite{c8thom2} (TR) have wrongly questioned the existence of such a gap of $\sim$ 1 eV in the electronic structure of graphite. They have also used double resonance using the electronic linear bands at K-point in the Brillouin zone. The serious difficulty with their calculations is that the calculated intensities of the D-mode for different incoming photon energies do not decrease as E$_{L}$ increases. On the contrary, the calculated intensity for E$_{L}$ = 4 eV is higher than for E$_{L}$ = 2 eV (see Fig.3 of Thomsen and Reich \cite{c8thom2}). This is completely opposite to the experiments as shown in Fig. 2.
There is another difficulty we have with the TR's calculation. As given by their Eq. (4), the magnitude of q increases with E$_L$. For optic branch along $\Gamma$K direction, the phonon frequency $\omega$ is a decreasing function of q and hence $\omega$ should decrease with increasing E$_L$ which is contrary to the experiments.

In this paper, we present a model based on double
resonance which can address all the novel features
associated with the D mode. Keeping in mind that the acoustic
phonon branches are strongly coupled to the high-frequency optic
branches at the K-point in graphite \cite{c8accop}, we propose that the D-mode is a
combination of an optic phonon at the K-point in the Brillouin zone
and an acoustic phonon whose momentum is determined uniquely by
"double resonance" condition \cite{c8card1}.  In usual second order scattering, if the
optic phonon is described by a wave-vector $\vec{q}_{1}$ and the
acoustic phonon by a wave-vector $\vec{q}_{2}$, conservation of
quasi-momentum requires $\vec{q}_{1}+\vec{q}_{2} =$ 0, where $\vec{q}_{1}$ and $\vec{q}_{2}$ can span the entire Brillouin zone. Since
D-mode is disorder-induced, we suggest that $\vec{q}_{1}$ is fixed at
the K-point of the Brillouin zone (maximum in the density of phonon states \cite{c8lesp1})
and the magnitude of $\vec{q}_{2}$ is determined by the double resonance condition \cite{c8card1}. The
quasi-momentum conservation is achieved by disorder-scattering. In a
double resonance process, real (not virtual) transition takes place
between two electronic states with emission of a phonon. Since the
electronic states have dispersion, phonons of appropriate wavevectors are
needed to keep the electronic transition real. Double resonance have been seen
for optic phonon overtones in GaAs \cite{c8diego1}, 2$\Gamma_{12}$ optic phonons
\cite{c8peter1} in Cu$_{2}$O and for two acoustic phonons  in
Ge \cite{c8bansal1}.

\section{Model}
Following Martin and Falicov \cite{c8card1}, we will discuss the two-phonon Raman scattering arising from an iterated one-phonon scattering in second order \cite{c8card1}. Near resonance condition, the incident photon is absorbed to create an electron-hole pair, with electron occupying the real
conduction band states and hole in the real valence state. Assuming parabolic dependence of the
electron-hole pairs in directions perpendicular to KH, the wave-vector of the electron or hole  $\vec{k}$  created by
a photon of energy E$_{L}$ can be written as 
$E_{L}=\Delta+\frac{{\hbar^{2}} k^{2}}{2 \mu}$
where $\mu$ is the reduced mass
$\left(\frac{1}{\mu}=\frac{1}{m_{e}^{\star}}+\frac{1}{m_{h}^{\star}}
\right)$, $m_{e}^{\star}$ and $m_{h}^{\star}$ are the electron and
hole effective masses and $\Delta$ is the band gap at K-point in the Brillioun zone.
We will now address the nature of critical point involved in the resonance
Raman scattering of D-band. 

We recall that the D-mode intensity
increases as  E$_{L}$ is decreased from 4.2 eV upto 1.16 eV, suggesting that the D-band excitation profile has a maximum below 1.2 eV. We suggest that the appropriate band gap associated with the resonance of the D-mode is
at $\sim$ 0.8 eV where a maximum  is seen
in the optical reflectivity measurements on graphite\cite{c8reflec,c8hanf1}. This gap is associated with the energy difference between E$_{1}$ and E$_{3}$ bands at the K point of the Brillouin zone as shown in the band structure of graphite (Fig.3a) taken from Ref.[19-21]. The wavevector is in the basal plane.
The bands shown here are the 4-$\pi$ bands in three-dimensional graphite.  Recall that a single sheet of graphite has 2-$\pi$ bands with linear dispersion and Fermi level passes through the point of interaction (K-point) as depicted in Fig.3b. Band structure calculations treating the interlayer coupling as a small perturbation find a small gap at K point in the Brillouin zone.
The optical transition between E$_{1}$ and E$_{3}$ is termed as A$_{2}$ transition. Within the Slonczewski-Weiss-McClure band model \cite{c8slon1} the energies of these optical transitions are mainly determined by the overlap interaction parameter $\gamma_{1} \sim$  0.4 eV for nearest-neighbour atoms on successive layers \cite{c8dress1} and energy of A$_2$ transition is $\sim$ 2$\gamma_{1}$. Thomsen and Reich \cite{c8thom2} have therefore, wrongly questioned the existence of a gap $\sim$ 0.8 eV in the 3D band structure of graphite as suggested Sood {\it et al.}\cite{c8sood1}

The mechanism we suggest is that the electron in the conduction band (or 
the hole in the valence band) is scattered by an optic phonon, changing the 
electron state from $k$ to $k^{\prime}$. Another accoustic phonon scatters 
this electron at $k^{\prime}$ to $k^{\prime\prime}$, followed by an impurity 
(disorder) scattering from $k^{\prime\prime}$ to $k^{\prime\prime\prime}$ and 
the electron-hole recombine to produce the scattered photon with a frequency 
shift given by $\omega = \omega_{op} + \omega_{ac}$. As mentioned before, we 
proposed that the optic phonon involved correspond to the maximum in the 
density of phonon states at the K-point in the Brillouin zone (shown by a 
solid dot in the phonon dispersion of graphite in Fig.3) and the wavevector 
q of the acoustic phonon is so chosen to satisfy the double resonance 
condition. This Raman process involving disorder mediated two-phonon 
scattering is fifth order in perturbation theory and will have four energy 
denominators as explicitly written by Kauschke {\it et al.} \cite{c8kaus1}, 
out of which two energy denominators will be the same. The resultant 
expression for the Raman matrix element is to be integrated over $k$, 
$k^{\prime}$, $k^{\prime\prime}$ and $k^{\prime\prime\prime}$. The Raman 
intensity will be maximum when all the three denominators resonate 
simultaneously. It can be shown \cite{c8card1,c8diego1,c8peter1,c8bansal1}
that the magnitude of q satisfy the 
condition for 
Stokes Raman scattering, for $\Delta + \hbar
\omega_{op}  <  E_{L} < \Delta + \hbar
\omega_{op}+ \hbar\omega_{ac}$, 
\begin{equation}
q_{s}=\left(\frac{\sqrt{2
\mu}}{\hbar}\right)\left[(E_{L}-\Delta)^{\frac{1}{2}} +
(E_{L}-\Delta-\hbar\omega_{op})^{\frac{1}{2}}\right]
\end{equation}
For   $E_{L} > \Delta + \hbar\omega_{op} + \hbar\omega_{ac}$, another
resonance becomes possible for
\begin{equation}
q_{s}=\left(\frac{\sqrt{2
\mu}}{\hbar}\right)\left[(E_{L}-\Delta-\hbar\omega_{op})^{\frac{1
} { 2}} +
(E_{L}-\Delta-\hbar\omega_{op}-\hbar\omega_{ac}(q))^{\frac{1}{2}}
\right]
\end{equation}
For a given E$_{L}$, the two-phonon Raman peak position for Stokes scattering will be given
by
\begin{equation}
\omega_{s} = \omega_{op}(K) + \omega_{ac}(q_{s})
\end{equation}
with $q_{s}$ given by Eq. (1) or (2). For $\Delta \sim$ 0.8 eV and E$_{L} >$ 1.16 eV, Eq.(2) is the appropriate relation to be used. In graphite, there is a peak in density of states around 1230 cm$^{-1}$ corresponding to the optic phonon frequency at the K point in the Brillioun zone \cite{c8accop}. We therefore, take   $\omega_{op}(K)$ to be 
1230 cm$^{-1}$. 

 Regarding acoustic phonon, it has been seen that the acoustic phonon branch corresponding to the branch marked T in the phonon dispersion relation of graphite shown in Fig.4, is strongly coupled to the optical phonon \cite{c8accop}. We take $\omega_{ac}(q) = v_{TA}q$ where $v_{TA}$ is the transverse accoustic phonon velocity (= 1.23$\times$10$^{4}$ m/s). Since
$\omega_{ac}\ll \omega_{op}$, one can simplify Eq.(2) and write
\begin{equation}
\omega_{ac}(q_{s}) = q_{s}v_{TA}=
2v_{TA}\left(\frac{\sqrt{2\mu}}{\hbar}\right)(E_{L}-\Delta-
\hbar\omega_{op})^{\frac{1}{2}}
\end{equation}

The solid line in Fig.1 is a fit to  Eq.(3) and (4), with
$\Delta$ and $\mu$ as adjustable parameters and $\omega_{op}$ = 1230 cm$^{-1}$. The
values obtained are $\Delta$ = 0.85 $\pm$ 0.05 eV and $\mu$ = 0.025 m$_e$. Taking the electron and hole effective mass to be same, $m_{e}^{\star} = 0.05m_{e}$, which is in excellent agreement with the values obtained from other experimental measurements like cyclotron resonance \cite{c8cyclo}. We also tried to fit the data in Fig.1 using Eq.(4), with $\omega_{op}$ also as an adjustable parameter along with $\Delta$ and $\mu$. In this case also the fit is very good with parameters, $\omega_{op}$ = 1211 cm$^{-1}$, $\Delta$ = 0.75 eV and $\mu$ = 0.026m$_{e}$. The values of the parameters in both cases being in resonable agreement with those obtained from other experiments gives us confidence in the double resonance model to understand the origin of D-band.

Recent Raman measurements on ion-implanted highly oriented pyrolytic
graphite by Tan {\it et al.}\cite{c8ptan1} show that the peak positions in the Stokes and anti-Stokes
spectra for the D band are different, with $\omega_{AS} > \omega_{S}$.  In our
model, using the disorder-induced double resonance, wavevector of the phonon in anti-Stokes Raman scattering for the D mode will be given by,
\begin{equation}
q_{AS}=\left(\frac{\sqrt{2
\mu}}{\hbar}\right)\left[(E_{L}-\Delta+\hbar\omega_{op})^{\frac{1
} { 2}} +
(E_{L}-\Delta+\hbar\omega_{op}+\hbar\omega_{ac}(q))^{\frac{1}{2}}
\right]
\end{equation}
in place of Eq.(2). Therefore
\begin{equation}
\Delta_{D} = \omega_{AS}-\omega_{S} = 2v_{TA}
\sqrt{\frac{2\mu}{\hbar^{2}}}\lbrace(E_{L}-\Delta+\omega_{op})^{\frac{1}{2}} -
(E_{L}-\Delta - \omega_{op})^{\frac{1}{2}}\rbrace
\end{equation}
For the D$^{\star}$ mode, two optic phonons and two acoustic phonons will be 
involved. This process will not require disorder-induced scattering to 
conserve momentum. In this case $\Delta_{D^{\star}} \sim 2\Delta_{D}$. 
Putting in the values for $\Delta$ = 0.85 eV and $\mu$ = 0.025m$_{e}$ as 
obtained from fit of Fig.1,  $\omega_{AS} - \omega_{S}$ for D mode = 13 cm$^{-1}$ (E$_{L}$ = 2.41 eV) while the difference observed experimentally \cite{c8ptan1} is $\sim$ 7 
cm$^{-1}$. In the case of
D$^{\star}$ mode we find the difference to be 26 cm$^{-1}$ which is very 
close to
the experimentally observed value of 25 cm$^{-1}$ of 
Tan {\it et al.} \cite{c8ptan1}.
Recent Stokes and anti-Stokes measurements by Zhang {\it et al.} \cite{c8zhan1} 
on carbon nanotubes also shows a  difference in the peak position of the D 
band in Stokes and anti-Stokes spectra and this difference is excitation 
laser energy dependent, very similar to the predictions of our model. 

In conclusion, we have addressed the long standing problem of the dispersion 
of the Raman D band as a function of laser excitation energy using a simple 
model based on disorder induced two phonon scattering and the double-resonance. The fit of the experimental data 
in this model yields the reduced effective 
mass of the electron-hole pair associated with the $\pi$ bands at the K-point. This is  in reasonable agreement with the 
values obtained from other experimental probes. As a corollary, this model 
can also explain the difference recently observed in the peak positions of the D and 
D$^{\star}$ bands in Stokes and anti-Stokes spectra.  

\section{acknowledgements}
AKS thanks Department of Science and Technology, India for financial 
assistance and Prof. Cardona for useful discussions. SAA acknowledges NIH grant GM30741 for financial support.
\begin{figure}
\label{Fig.1}
\caption{D-mode frequency dependence on excitation energy for
different forms of carbon-disordered carbon \cite{c8wang1} (filled squares) , glassy carbon \cite{c8pocsik1} (filled triangles) and
carbon nanotubes \cite{c8thomsen1} (open circles).}
\end{figure}

\begin{figure}
\label{Fig.2}
\caption{Raman spectra of graphite at E$_{L}$ = 4.82 and 2.54 eV. The inset shows the intensity ratio of D-band to that of G-band as a function of E$_{L}$ as reported by Wang {\it et al.} \cite{c8wang1}. The peaks marked by asteriks are the internal vibrational modes of sulphate ion for use as internal intensity calibrant. 
}
\end{figure}

\begin{figure}
\label{Fig.3}
\caption{(a) Schematic band structure of graphite showing the four $\pi$ bands at the K-point of the Brillouin zone. The wavevector {\bf k} is perpendicular to the c-axis. (b) Schematic of the single graphene sheet at the K-point. Dashed lines denote the Fermi energy.}
\end{figure}

\begin{figure}
\label{Fig.4}
\caption{Phonon dispersion curves in graphite calculated along certain high symmetry axes as given by Mathew {\it et al.}\cite{c8mathew1}.
}
\end{figure}

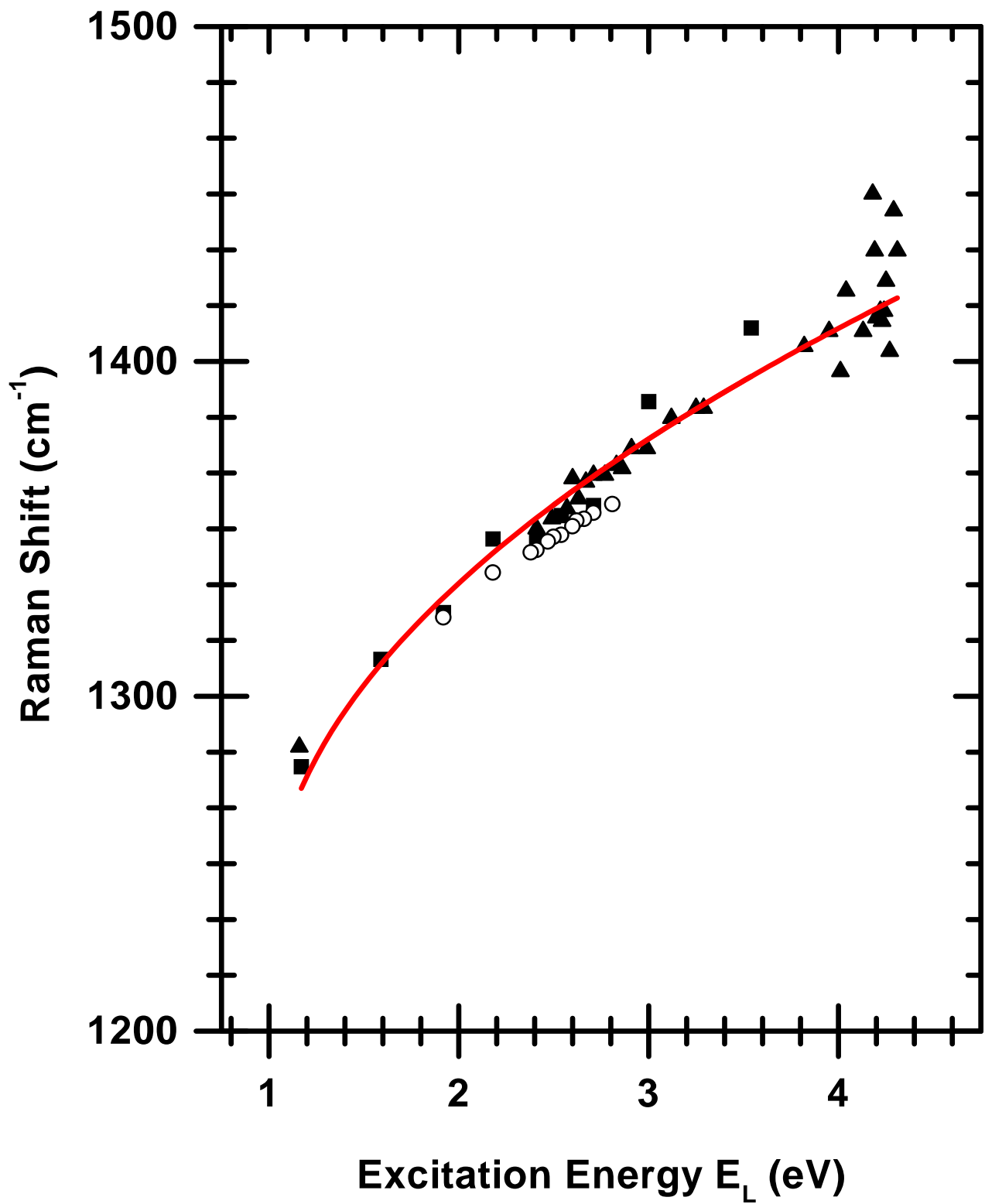
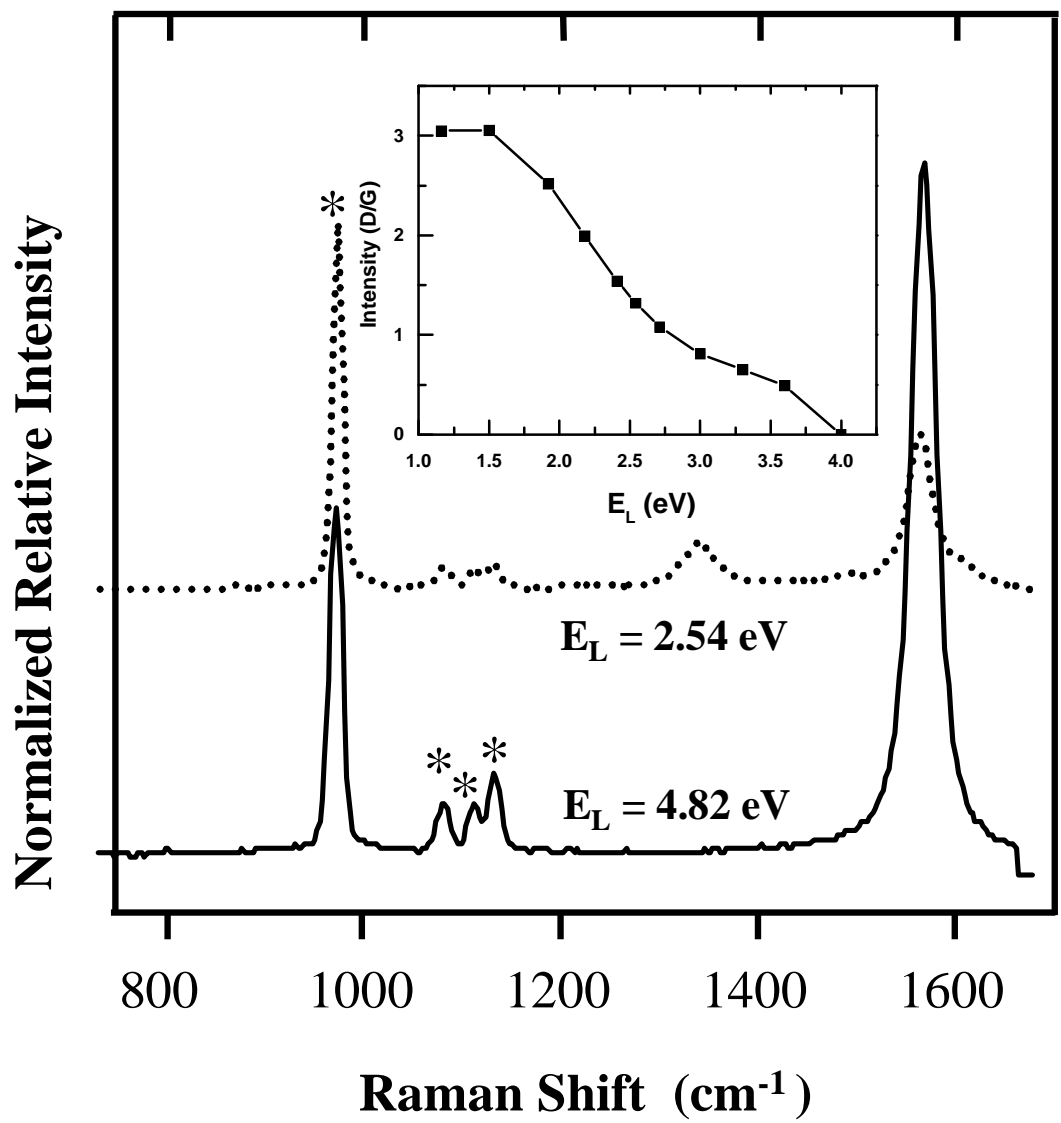
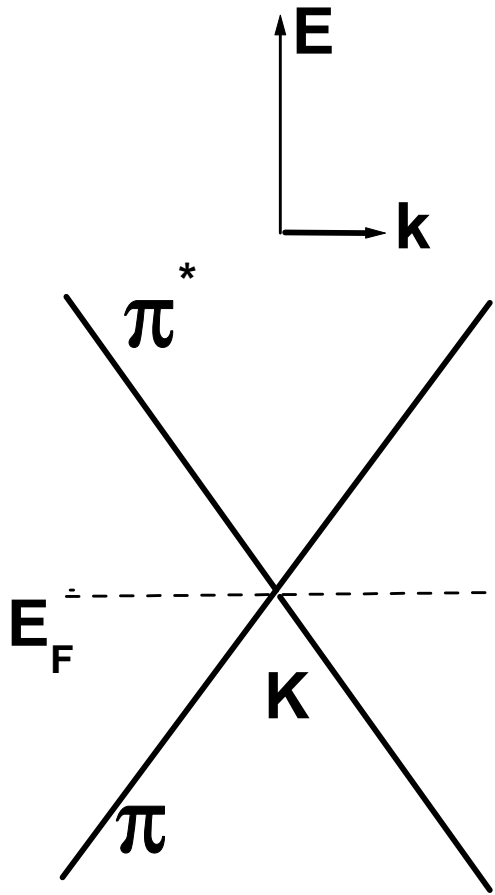
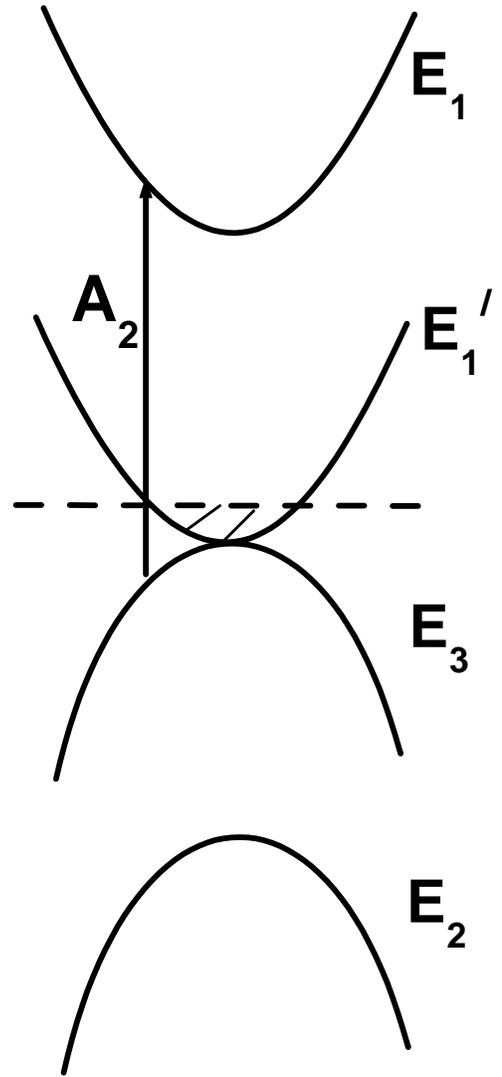
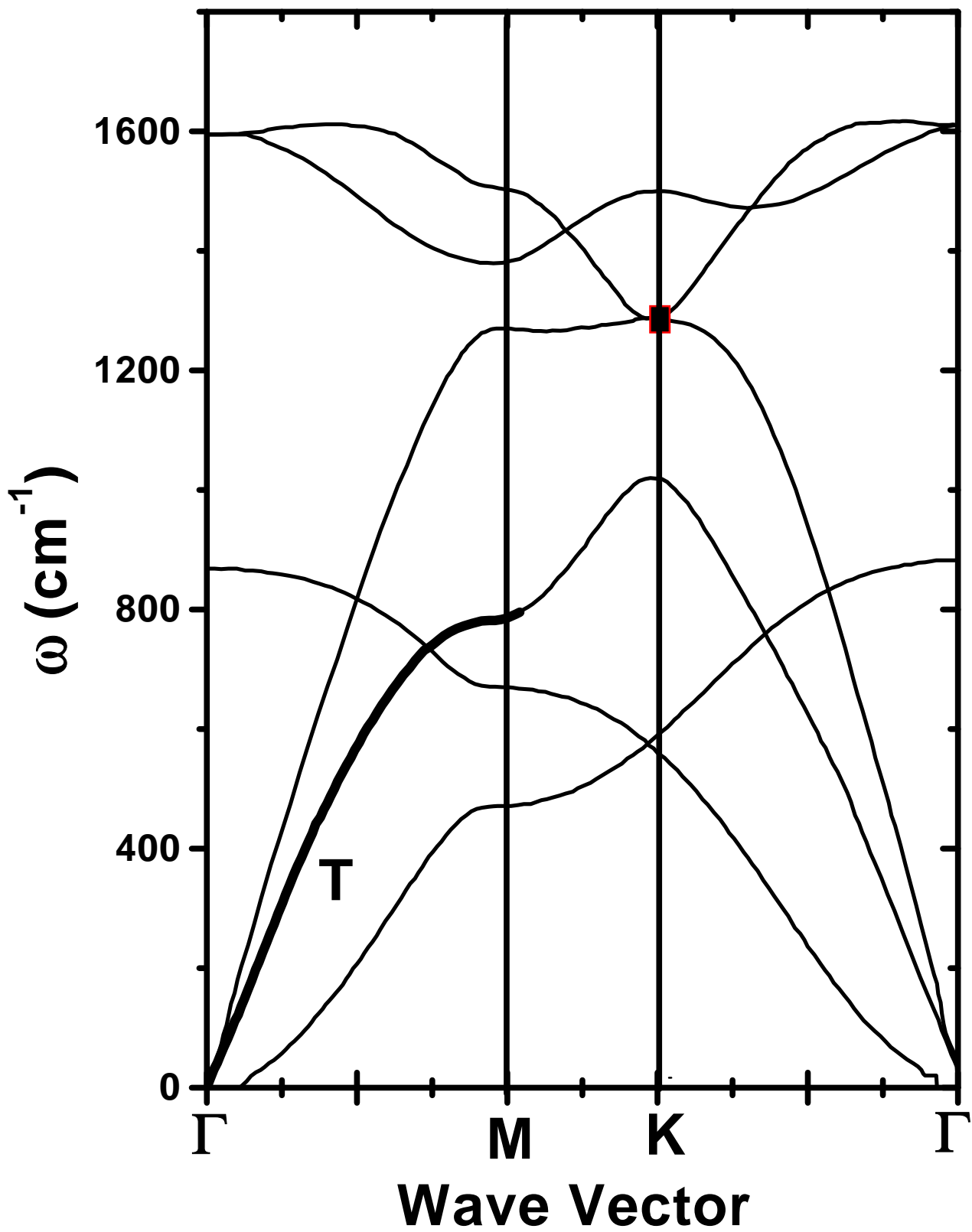
\end{document}